\documentclass[manuscript,screen]{acmart}

\AtBeginDocument{%
  \providecommand\BibTeX{{%
    \normalfont B\kern-0.5em{\scshape i\kern-0.25em b}\kern-0.8em\TeX}}}

\setcopyright{acmcopyright}
\acmJournal{TOCHI}
\acmYear{2022} \acmVolume{1} \acmNumber{1} \acmArticle{1} \acmMonth{1} \acmPrice{15.00}\acmDOI{}

\begin{document}

\title{Designing Creative AI Partners with COFI: A Framework for Modeling Interaction in Human-AI Co-Creative Systems}

\author{Jeba Rezwana}
\email{jrezwana@uncc.edu}
\author{Mary Lou Maher}
\email{m.maher@uncc.edu}
\affiliation{%
  \institution{University of North Carolina at Charlotte}
  \country{USA}
  \postcode{}
}
\renewcommand{\shortauthors}{Rezwana and Maher}

\begin{abstract}
Human-AI co-creativity involves both humans and AI collaborating on a shared creative product as partners. In a creative collaboration, interaction dynamics, such as turn-taking, contribution type, and communication, are the driving forces of the co-creative process. Therefore the interaction model is a critical and essential component for effective co-creative systems. There is relatively little research about interaction design in the co-creativity field, which is reflected in a lack of focus on interaction design in many existing co-creative systems. The primary focus of co-creativity research has been on the abilities of the AI. This paper focuses on the importance of interaction design in co-creative systems with the development of the Co-Creative Framework for Interaction design (COFI) that describes the broad scope of possibilities for interaction design in co-creative systems. Researchers can use COFI for modeling interaction in co-creative systems by exploring alternatives in this design space of interaction. COFI can also be beneficial while investigating and interpreting the interaction design of existing co-creative systems. We coded a dataset of existing 92 co-creative systems using COFI and analyzed the data to show how COFI provides a basis to categorize the interaction models of existing co-creative systems. We identify opportunities to shift the focus of interaction models in co-creativity to enable more communication between the user and AI leading to human-AI partnerships.
\end{abstract}



\maketitle

\section{Introduction}
Computational creativity is an interdisciplinary field that applies artificial intelligence to develop computational systems capable of producing creative artifacts, ideas and performances \cite{colton2012computational}. Research in computational creativity has lead to different types of creative systems that can be categorized based on their purposes: systems that generate novel and valuable creative products, systems that support human creativity, and systems that collaborate with the user on a shared creative product combining the creative ability of both the user and the AI \cite{davis2015enactive}. Davis introduces the term human-computer co-creativity, where humans and computers can collaborate in a creative process as colleagues \cite{davis2013human}. In human-computer co-creativity, both humans and AI agents are viewed as one system through which creativity emerges. The creativity that emerges from a collaboration is different from creativity emerging from an individual as creative collaboration involves interaction among collaborators and the shared creative product is more creative than each individual could achieve alone \cite{sawyer2009distributed}. Stephen Sonnenburg demonstrated that communication is the driving force of collaborative creativity \cite{sonnenberg1991strategies}. Interaction is a basic and essential component of co-creative systems as both the human and the AI actively participate and interact in the co-creation, unlike autonomous creative systems that generate creative artifacts alone and creativity support tools that support human creativity. 

Designing and evaluating co-creative systems has many challenges due to the open-ended and improvisational nature of the interaction between the human and the AI agent \cite{davis2016empirically, kantosalo2014isolation}. Humans utilize many different creative strategies and reasoning processes throughout the creative process, and ideas and the creative product develop dynamically through time. This continual progression of ideas requires adaptability on the agent's part. Additionally, it is not always clear how the co-creative AI should contribute and interact during the course of the co-creative process. For example, sometimes the human may want to lead and have the AI assist with some tasks, whereas other times the human may want the AI to lead to help find inspiration or to work independently. Understanding the mechanics of co-creation is still very much open questions in the young field of human-computer co-creativity. Bown asserted that the success of a creative system's collaborative role should be further investigated as interaction plays a key role in the creative process of co-creative systems \cite{bown2015player}. AI ability alone does not ensure a positive collaborative experience of users with the AI \cite{louie2020novice} and interaction is more critical than algorithms where interaction with the users is essential \cite{wegner1997interaction}. In this paper we focus on the interaction design space as an essential aspect of effective co-creative systems.

Interaction design is the creation of a dialogue between users and the system \cite{kolko2010thoughts}. Recently, interaction design in co-creative systems is being addressed as a significant aspect of computational creativity. Kantosalo et al. said that interaction design, specifically, interaction modality should be the ground zero for designing co-creative systems \cite{kantosalo2020modalities}. However, the interaction designs of many existing co-creative systems provide only one-way interaction, where humans can interact with the AI but the system is not designed for the AI to communicate back to humans. For example, Collabdraw \cite{fan2019collabdraw} is a co-creative sketching environment where users draw with an AI. The user interface includes only one button that users click to submit their artwork and indicate that their turn is complete. Other than the button, there is no way for the user to communicate with the AI and vice versa to provide information, suggestions, or feedback. Although the AI algorithm is capable of providing intriguing contributions to the creative process, the interaction design is inadequate for collaboration between human and AI. Another example, Image to Image \cite{isola2017image} is a co-creative system that converts a line drawing of a particular object from the user into a photo-realistic image. The user interface has only one button that users click to tell the AI to convert the drawing. Interaction design can provide more than the transfer of instructions from a user to an AI agent to generate a creative artifact and can lead to a more engaging user experience. A recent study showed increased user satisfaction with text-based instructions from the AI rather than button-based instructions in a co-creation \cite{oh2018lead}. A starting point to investigate interaction models is the study of collaboration among humans \cite{davis2015enactive}. Understanding the factors in human collaboration can build the foundation for the development of human-AI collaboration in co-creative systems \cite{mamykina2002collaborative}. Interaction models developed for computer supported collaborative work is an important source for identifying interaction models related to co-creative systems. 

In this paper, we present Co-Creative Framework for Interaction design (COFI) that describes interaction components as a space of possibilities for interaction design in co-creative systems. These interaction components represent various aspects of a co-creation, such as participation style, contribution type, and communication between humans and the AI. COFI is informed by the literature on human collaboration, CSCW, computational creativity, and human-computer co-creativity. We adopted interaction components based on a literature review and adapted the components to concepts relevant to co-creativity. COFI can be used as a guide when designing the interaction models in co-creative systems. COFI can also be beneficial for investigating and interpreting the interaction design of existing co-creative systems. We coded and analyzed the interaction models of a dataset of 92 co-creative systems using COFI to evaluate the value and analytical power of the framework. Three distinct interaction models for co-creative systems emerged from this analysis: generative pleasing AI agents that follow along with the user, improvisational AI agents that work alongside users on a shared product spontaneously, and advisory AI agents that both generate and evaluate the creative product. The analysis reveals that the co-creative systems in this dataset lack communication channels between the user and AI agent. Finally, this paper discusses the limitations in the existing interaction models in co-creative systems, potential areas for further development, and the importance of extending the scope of human-AI communication in co-creative systems. 

\vspace{-0.3cm}  
\section{Related Work}

\subsection{Co-creative Systems}
Creativity is defined as the exploration and production of novel and useful ideas \cite{dietrich2004cognitive, jung2013structure, jennings2011understanding}. Wiggins defined creative systems as "A collection of processes, natural or automatic, which are capable of achieving or simulating behavior which in humans would be deemed creative \cite{wiggins2006preliminary}." Davis et al. discussed the three main categories of creative systems based on their working processes and purposes \cite{davis2015enactive}: standalone generative systems, creativity support tools, and co-creative systems. Standalone generative systems refer to fully autonomous intelligent systems that work independently without any interaction with humans in the creative process. Creative systems that support the user's creativity  without contributing to the creative process are considered creativity support tools (CST). In co-creative systems, humans and computer both contribute as creative colleagues in the creative process\cite{davis2013human}. Co-creative systems originated from the concept of combining standalone generative systems with creativity support tools as computers and humans both take the initiative in the creative process and interact as co-creators \cite{kantosalo2020role}. Mixed initiative creative systems are often used as a substitute term for co-creative systems in the literature \cite{yannakakis2014mixed}. 

In a co-creative system, interaction between the human and AI agent make the creative process complex and emergent. Maher explores issues related to \emph{who} is being creative when humans and AI collaborate in a co-creative system \cite{maher2012computational}. Antonios Liapis et al. argued that when creativity emerges from human-computer interaction, it cannot be credited either to the human or to the computer alone, and surpasses both contributors' original intentions as novel ideas arise in the process \cite{liapis2014computational}. Designing interaction in co-creative systems has unique challenges due to the spontaneity of the interaction between the human and the AI \cite{davis2016empirically}. A co-creative AI agent needs continual adjustment and adaptation to cope with human strategies. A good starting point to investigate questions about modeling an effective interaction design for co-creative systems can be studying creative collaboration in humans \cite{davis2015enactive}. Mamykina et al. argued that by understanding the factors of human collaborative creativity, methods can be devised to build the foundation for the development of computer-based systems that can augment or enhance collaborative creativity \cite{mamykina2002collaborative}. 
\vspace{-0.3cm}                  

\subsection{Interaction Design in Co-creative Systems}
 Regarding interaction design in interactive artifacts, Fallman stated: "interaction design takes a holistic view of the relationship between designed artifacts, those that are exposed to these artifacts, and the socio-cultural context in which the meeting takes place" \cite{fallman2008interaction}. In the field of co-creativity, interaction design includes various parts and pieces of the interaction dynamics between the human and the AI, for example - participation style, communication between collaborators, and contribution type. Now the question is how researchers and designers can explore the possible spaces of interaction design in co-creative systems. For instance, turn-taking is the ability for agents to lead or follow in the process of interaction \cite{winston2017turn}. While designing a co-creative system, should the designer consider turn-taking or concurrent participation style? Turn-taking models work well in many co-creative systems but may not fit well for all co-creative systems. Lauren and Magerko investigated whether the user experience is improved with a turn-taking model applied to Lumin AI, a co-creative dance partner, through an empirical study \cite{winston2017turn}. However, their results showed negative user experience with a turn-taking model compared to a non-turn taking model. The negative user experience resulted from the dislike for the AI agent to take the lead. 

Bown argued that the most practiced form of evaluating artificial creative systems is mostly theoretical and is not empirically well-grounded and suggested interaction design as a way to ground empirical evaluations of computational creativity \cite{bown2014empirically}. Yee-King and d'Inverno also argued for a stronger focus on the user experiences of creative systems, suggesting a need for further integration of interaction design practice into co-creativity research \cite{yee2016experience}. There is a lack of a holistic framework for interaction design in co-creative systems. A framework for interaction design is necessary to explain and explore the possible interaction spaces and compare and evaluate the interaction design of existing co-creative systems for improving the practice of interaction modeling in co-creative systems.

 There are recent developments in frameworks and strategies for interaction in co-creative systems. Kantosalo et al. proposed a framework to describe three aspects of interaction, interaction modalities, interaction style and interaction strategies, in co-creative systems \cite{kantosalo2020modalities}. They analyzed nine co-creative systems with their framework to compare different systems' creativity approaches even if they are within the same creative domain \cite{kantosalo2020modalities}. Bown and Brown identified three interaction strategies - operation-based interaction, request-based interaction and ambient interaction in metacreation, the automation of creative tasks with machines \cite{bown2018interaction}. Bown et al. explored the role of dialogue between the human and the user in co-creation and argued that both linguistic and non-linguistic dialogues of concepts and artifacts are essential to maintain the quality of co-creation \cite{bown2020speculative}. Guzdial and Riedl proposed an interaction framework for turn-based co-creative AI agents to better understand the space of possible designs of co-creative systems \cite{guzdial2019interaction}. Their framework is limited to turn-based co-creative agents and has a focus on contributions and turn-taking. In this paper we present COFI, a description of a design space of possibilities for interaction in co-creative systems that includes and extends these existing frameworks and strategies.

\vspace{-0.2cm}  
\subsection{Creative Collaboration among Humans}
Sawyer asserted that the creativity that emerges from collaboration is different from the creativity emerging from an individual where interaction among the group is a vital component of creativity \cite{sawyer2009distributed}. He investigated the process of creativity when emerging from a group by observing and analyzing improvisational theater performances by a theater group \cite{sawyer2009distributed} and argued that the shared product of collaborative creativity is more creative than each individual alone could achieve. Sonnenburg introduced a theoretical model for creative collaboration, and this model presents communication among the group as the driving force of collaborative creativity \cite{sonnenberg1991strategies}. Interaction among the individuals in a collaboration makes the process emergent and complex. For investigating human collaboration, many researchers stressed the importance of understanding the process of interaction. Fantasia et al. proposed an embodied approach of collaboration which considers collaboration as a property and intrinsic part of interaction processes \cite{fantasia2014we}. They claimed that interaction dynamics help in understanding and fostering our knowledge of different ways of engaging with others and argued that it is crucial to investigate the interaction context, the environment, and how collaborators make sense of the whole process for gaining more knowledge and understanding more about collaboration. In COFI, we include components that address the interaction we observe in human to human collaboration as possibilities for human-AI co-creativity.

Computer supported cooperative work (CSCW) is computer assisted coordinated activity carried out by a group of collaborating individuals \cite{baecker1993readings}. K Schmidt defined CSCW as an endeavor to understand the nature and characteristics of collaborative work to design adequate computer-based technologies \cite{schmidt2008cooperative}. A foundation of CSCW is sense-making and understanding the nature of collaborative work for designing adequate computer based technology to support human collaboration. CSCW systems are designed to improve group communication while alleviating negative interactions that reduce collaboration quality \cite{kamel1998applying}. For building effective CSCW systems for collaborative creative work, many CSCW researchers investigated creative collaboration among humans to understand the mechanics of collaboration. For this reason, the design space of CSCW is relevant for interaction design in co-creative systems.

\vspace{-0.4cm}  

\subsection{Sense-making in Collaboration}
Sense-making is motivated by a continual urge to understand connections among people, places, and events to anticipate their trajectories \cite{klein2006making}. Russel et al. discussed the processes involved in a sense-making: searching for representations, instantiate representations by encoding the representations, and utilizing the encoding in task-specific information \cite{russell1993cost}. Davis argued that participatory sense-making is useful to analyze, understand and model creative collaboration \cite{davis2015enactive}. De Jaegher and Di Paolo also proposed participatory sense-making as a starting point of understanding social interaction \cite{de2007participatory}. To understand participatory sense-making, the definition of sense-making from cognitive theory is crucial. Sense-making is the way cognitive agents meaningfully connect with their world, based on their needs and goals as self-organizing, self-maintaining, embodied agents \cite{de2013embodiment}. Introducing multiple agents in the environment makes the dynamics of sense-making more complex and emergent as each agent is interacting with the environment as well as with each other. Participatory sense-making evolves from this complex, mutually interactive process \cite{davis2015enactive}. Participatory sense-making occurs where
"A co-regulated coupling exists between at least two autonomous agents where the regulation itself is aimed at the aspects of the coupling itself so that the domain of relational dynamics constitutes an emergent autonomous organization without destroying the autonomy of the agents involved." \cite{de2007participatory}. In this quote, De Jaegher and Di Paolo outlines the process of participatory sense-making where meaning-making of relational interaction dynamics such as the rhythm of turn-taking, manner of action, interaction style, etc is necessary \cite{davis2016empirically}. 

\begin{figure}[h]
 \vspace{-0.1cm}
  \centering
  \includegraphics[width=0.55\linewidth]{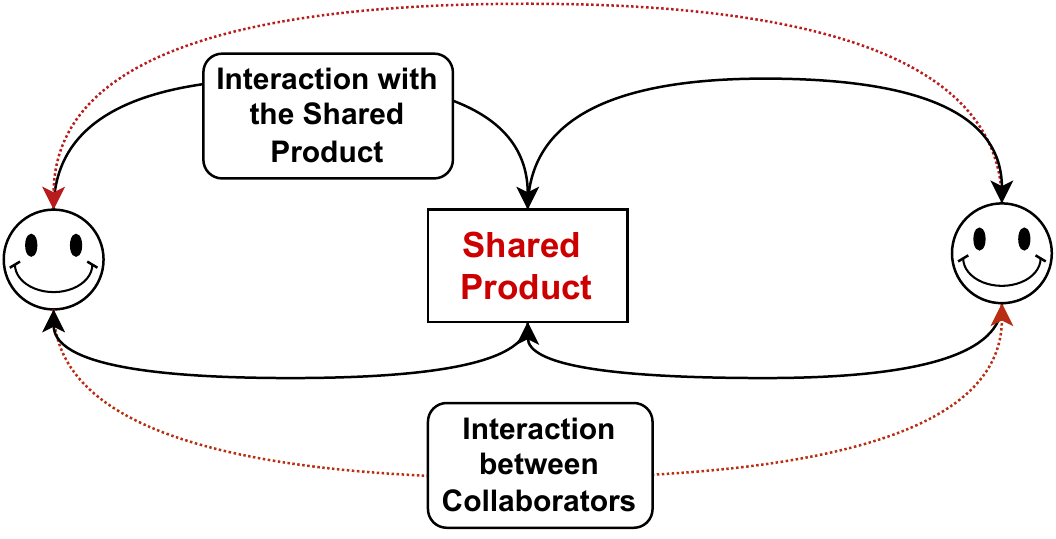}
  \caption{Interactional Sense-making in a Co-creation.}
  \Description{}
  \vspace{-0.1cm}
\end{figure}

To understand interaction dynamics in an open-ended improvisational collaboration, Kellas and Trees present a model of interactional sense-making \cite{kellas2005rating}. They describe two types of interaction in the sense-making process: interaction between collaborators and interaction with the shared product (Figure 1). We adapt and extend this model for COFI to ground our space of possibilities for interaction design on the concept of interactional sense-making. Interaction with the shared product, in the context of a co-creative system, describes the ways in which the co-creator can sense, contribute, and edit content of the emerging creative product. Interaction between collaborators explains how the interaction between the co-creators is unfolding through time which includes turn-taking, timing of initiative, communication etc. Participatory sense-making occurs when there is a mutual co-regulation of these two interactional sense-making processes between the co-creators \cite{}. For example, when both participants are adapting their responses based on each other’s contribution while maintaining an engaging interaction dynamic, participatory sense-making occurs.

\section{Co-Creative Framework for Interaction Design (COFI)}
We develop and present Co-Creative Framework for Interaction Design (COFI) as a space of possibilities for interaction design in co-creative systems. COFI also provides a framework for analyzing the interaction design trends of existing co-creative systems. This framework describes various aspects involved in the interaction between the human and the AI. COFI is informed by research on human collaboration, CSCW, computational creativity, and human-computer co-creativity. 

\begin{figure}[h]
  \centering
  \includegraphics[width=16cm, height=11cm]{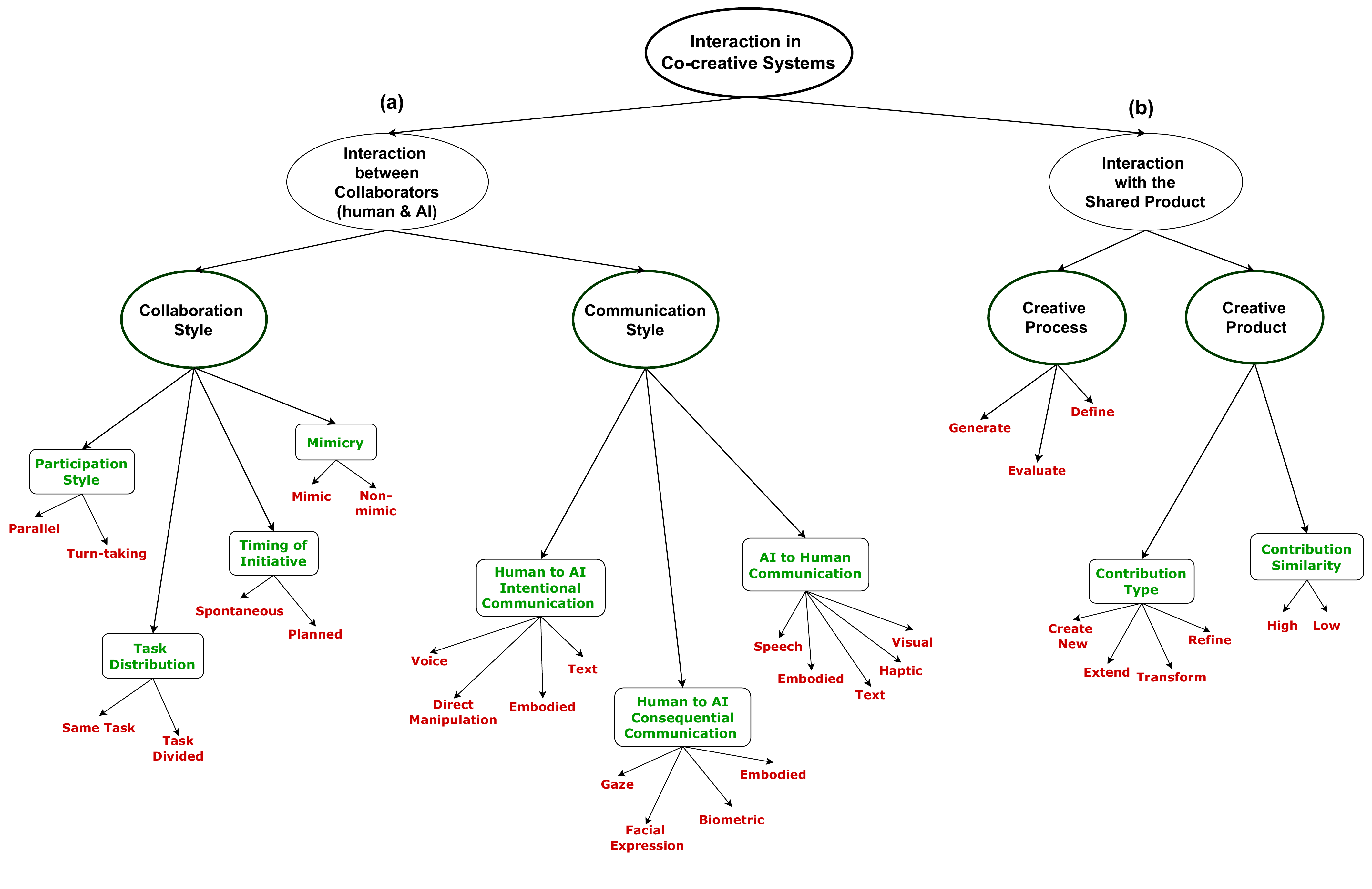}
  \caption{Co-Creative Framework for Interaction Design (COFI): On the left (a) Components of Interaction between the collaborators, On the right (b) Components of Interaction with the Shared Product.}
  \Description{}
  \label{COFI}
  \vspace{-0.4cm}
\end{figure}

The primary categories of COFI are based on two types of interactional sense-making of collaboration as described by Kellas and Trees (Figure \ref{COFI}) \cite{kellas2005rating}: interaction between collaborators and interaction with the shared product. Interaction with the shared product, in the context of co-creative systems, describes interaction aspects related to the creation of the creative content. Interaction between collaborators explains how the interaction between the human and the AI is unfolding through time which includes turn-taking, timing of initiative, communication, etc. Thus, COFI characterizes relational interaction dynamics between the collaborators (human and AI) as well as functional aspects of interacting with the shared creative product. Kellas and Trees' framework was used for explaining and evaluating the interaction dynamics in human creative collaboration in joint storytelling. Understanding collaborative creativity among humans can be the basis for designing effective co-creative systems where the AI agent acts as a creative partner. 

Each of the two categories of interaction is further divided into two subcategories. Interaction between collaborators is divided into collaboration style and communication style. On the other hand, interaction with the shared product is divided into the creative process and creative product. CSCW literature discusses collaboration mechanics among the collaborators to make effective CSCW systems. Many frameworks about groupware and CSCW systems discuss and emphasize both collaboration components and communication components among collaborators. For example, Baker et al. proposed an evaluation technique based on collaboration mechanics for groupware and emphasized both coordination and communication components in a collaboration \cite{baker2001heuristic}. Creativity literature focuses more on creativity emergence, which includes creative processes and the creative product. For example, Rhodes's famous 4P, which is one of the most acknowledged model, includes creative process and product \cite{rhodes1961analysis}. Therefore, in COFI, the literature regarding human collaboration and CSCW literature informs the category 'interaction between the collaborators', while the creativity and co-creativity literature provides descriptions of the 'interaction with the shared product'. In human-AI co-creativity, the focus should be on both creativity and collaboration. As a result, both the CSCW and creativity literature provide the basis for defining the interaction components of COFI under the four subcategories. 

We performed a literature review to identify the components of COFI. We identified a list of search databases for relevant academic publications: ACM Library, arXiv, Elsevier, Springer, and ScienceDirect, and google scholar. We used keywords based on the 4 Cs in COFI: Collaboration style, Communication style, Creative process, Creative product. The total list of keywords are: 'human collaboration mechanics,' 'creative collaboration among humans,' 'communication in collaboration,' 'cooperation mechanics,' 'Interaction in joint action,' 'groupware communication,' 'interaction design in computational creativity,' 'interaction in co-creativity,' 'creative process,' 'group interaction in computational creativity,' 'interaction in human-computer co-creation'. We considered documents published from 1990 until 2021. We did not include papers that are a tutorial or poster, papers that are not in English, papers that by title or abstract are outside the scope of the research, and papers that do not describe the collaboration mechanics or group interaction. We included papers describing strategies, mechanisms and components of interaction in a natural collaboration, computer-mediated collaboration and human-AI collaboration. COFI was developed in an iterative process of adding, merging, and removing components based on the interaction components defined in the literature. We refer to the specific publications that contributed to each component of COFI in the sections below: for each interaction component the first paragraph defines the component, and the second paragraph references the relevant publications that provided the basis for that component.

\subsection{Interaction between Collaborators (Human and AI)}
This section presents components related to the relational interaction dynamics between the human and the AI as co-creators. As shown in Figure 1(a), interaction between collaborators is divided into two subcategories which are \emph{collaboration style} and \emph{communication style}.

\subsubsection{Collaboration Style}
Collaboration style is the manner of working together in a co-creation. In COFI, collaboration style comprises participation style, task distribution, timing of initiative and mimicry as interaction components. The following subsections describe each interaction component in this category.

\textbf{Participation Style:}
Participation style in COFI refers to whether the collaborators can participate and contribute simultaneously, or one collaborator has to wait until the partner finishes a turn. Therefore, participation style in COFI is categorized as parallel and turn-taking. For example, in a human-AI drawing co-creation, collaborators can take turns to contribute to the final drawing or they can draw simultaneously.

Participation style in COFI is based on the categorization of interpersonal interaction into two types: concurrent interaction and turn-based interaction \cite{liu2015role}. In concurrent interaction, continuous parallel participation from the collaborators occurs and in turn-based interaction, participants take turns in contributing. In a parallel participation style, both collaborators can contribute and interact simultaneously \cite{penichet2007classification}. In a turn-taking setting, simultaneous contribution can not occur \cite{penichet2007classification}. In CSCW research, there is a concept for interaction referred to as synchronous and asynchronous. Synchronous interaction requires the real time interaction where the presence of all collaborators is required. Whereas asynchronous cooperation does not require simultaneous interaction of all collaborators \cite{rodden1991cscw, reinhard1994cscw, cacciagrano2001synchronous}. In CSCW, the distinction between synchronous and asynchronous interaction is information exchange in terms of time. In COFI, participation style describes the way collaborators participate when all are present at the same time.

\textbf{Task Distribution:}
Task distribution refers to the distribution of tasks among the collaborators in a co-creative system. In COFI, there are two types of task distribution, same task and task divided. When it is same task, there is no division of tasks between collaborators and all the collaborators take part in the same task. For example, in a human-AI co-creative drawing, both co-creators do the same task, i.e. generating the drawing. In a task-divided distribution, the main task is divided into specific sub-tasks and the sub-tasks are distributed among the collaborators. For example, in co-creative poetry, the user can define the conceptual space for the poetry and generate a poem while the AI agent can evaluate the poetry. 

Cahan and Fewell asserted that division of task is a key factor in the success of social groups \cite{cahan2004division}. According to Fischer and Mandl, task division should be addressed for co-ordination in a computer-mediated collaboration \cite{fischer2003being}. This component of COFI emerged from discussions of the two interaction modes presented by Kantosalo and Toivonen: alternating co-creativity and task divided co-creativity \cite{kantosalo2016modes}. In alternating co-creativity, each party contributes to the shared artifact while doing the same task by taking turns. Kantosalo and Toivonen emphasized the turn-taking in alternating interaction mode. In COFI, we renamed alternating co-creativity to be same task as we want to emphasize the task distribution. Task divided in COFI is the same term used in Kantosalo and Toivenen \cite{kantosalo2016modes}.

\textbf{Timing of Initiative:}
In a co-creative setting, the timing of collaborators' initiative can be scheduled beforehand, or it can be spontaneous. If the timing of the initiative is planned or fixed in advance, in COFI it will be addressed as planned. If both agents initiate their contribution without any prior plan or fixed rules, it will be addressed as spontaneous. Timing of the initiative should be chosen based on the motivation behind designing a co-creative system. Spontaneous timing is suitable for increased emergent results, whereas planned timing is more suitable for systems where users want inspiration or help in a specific way for a particular aspect of the creative process. 

Salvador et al. discussed timing of initiative in their framework for evaluating groupware for supporting collaboration \cite{salvador1996denver}. They defined two types of timing of initiative: spontaneous initiatives, where participants take initiatives spontaneously and pre-planned initiatives, where group interactions are scheduled in advance. Alam et. al divided interaction among groups into planned and impromptu \cite{alam2013computer}. For COFI, we merged these ways of describing the timing of initiative into spontaneous and planned.

\textbf{Mimicry:}
COFI includes mimicry as a subcategory of collaboration style which is used in co-creative systems as an intentional strategy for collaboration. When mimicry is a strategy for the AI contribution, the co-creative AI mimics the human user.

Drawing Apprentice \cite{davis2015drawing} is a co-creative web-based drawing system that collaborates with users in real-time abstract drawing while mimicking users. The authors demonstrated with their findings that even if the Drawing Apprentice mimics the user in the creative process, the system engaged users in the creative process that resulted in generating novel ideas. An example of a non-mimic co-creative system is Viewpoints AI. Viewpoints AI is a co-creative system where a human can engage in collaborative dance movement as the system reads and interprets the movement for responding with an improvised movement \cite{jacob2013viewpoints}. 

\subsubsection{Communication Style}
In COFI, communication style refers to the ways humans and AI can communicate. Communication is an essential component in any collaboration for the co-regulation between the collaborators and helps the AI agent make decisions in a creative process \cite{bown2020speculative}. Communication is critical for achieving understanding and coordination between collaborators. A significant challenge in human-AI collaboration is the development of common ground for communication between humans and machines \cite{dafoe2021cooperative}. Collaborators communicate in different ways in a co-creation such as, communication through the shared product and contributions, and communication through different communication channels or modalities. In co-creative systems, collaborators contribute to the shared product through the creative process and sense-making of each others' contributions during the process and act accordingly. Communicating through the shared product is a prerequisite in a co-creation or any collaborative system \cite{bown2018interaction}. Hence, COFI does not include interaction through the shared product under \emph{communication style}. In COFI, \emph{communication style} includes different channels or modalities designed to convey intentional and unintentional information between users and the AI. Human to AI communication channels carry information from users to the AI. On the other hand, AI to human communication channels carry information from the AI to users. 

\textbf{Human to AI Intentional Communication:}
Human to AI intentional communication channels represent the possible ways a human agent can intentionally and purposefully communicate to the AI agent to provide feedback and convey important information. In COFI, human to AI communication channel includes direct manipulation, voice, text and embodied communication. The human agent can directly manipulate the co-creative system by clicking buttons for giving instructions, feedback, or input. It can also provide user preferences by selecting from AI provided options. Using the whole body or gestures for communicating with the computer will be referred to as embodied. Voice and text can be also used as intentional communication channels from human to AI. 

Gutwin and Greenburg proposed a framework that discusses the mechanics of collaboration for groupware \cite{gutwin1996workspace}. Their framework includes seven major elements and one of them is explicit or intentional communication. Bard defined intentional communication as the ability to coordinate behavior involving agents \cite{bard1992intentional}. Brink argued that the primary goal of intentional communication is to establish joint attention \cite{brinck2008role}. In the field of human-computer interaction, the communication channel between humans and computers is described as a modality. The modalities for intentional communication from human to AI include direct manipulation, embodied/gesture, text, and voice \cite{nigay2004design}.

\textbf{Human to AI Consequential Communication:}
In COFI, human to AI consequential communication channels represent the ways the human user unintentionally or unconsciously gives off information to the AI agent. In other words, this channel represents the ways a co-creative AI agent can track and collect unintentional or consequential information from the human user such as eye tracking, facial expression tracking, biometric data tracking and embodied movements. AI agents can track and collect various consequential details from the human to perceive user preference, user agency and engagement. For example, a posture or facial expression can indicate boredom or lack of interest.

Gutwin and Greenburg reported consequential or unintentional communication as a major element of collaboration mechanics, in addition to intentional communication \cite{gutwin1996workspace}. Collaborators pick up important information that is unintentionally "given off'' by others, which is considered as consequential communication in a human collaboration. Unintentional communication, such as embodied communication, gaze, biometric measurement and facial expression are consequential communication \cite{gutwin1996workspace}. Revealing the internal state of an individual is termed 'Nonverbal leakage' by Ekman and Freisen \cite{ekman1969nonverbal}. Mutlu et al. argued that in a human-AI interaction, unintentional cues have a significant impact on user experience \cite{mutlu2009nonverbal}. 

\textbf{AI to Human Communication:}
AI to human communication represents the channels through which AI can communicate to humans. Humans expect feedback, critique and evaluation of our contribution from collaborators in teamwork. If the AI agent could communicate their status, opinion, critique and feedback for a specific contribution, it would make the co-creation more balanced as the computational agent will be perceived as an intelligent entity and a co-equal creative partner rather than a mere tool. This communication involves intentional information from the AI to human. Because the interaction abilities of a co-creative AI agent are programmed, all of the communication from the AI is intentional. However, one may ask, can AI do anything unintentional or unconscious beyond the programmed interaction? A co-creative AI can have a body and can make a facial expression of boredom. However, can we call it unintentional or it is also an intentional information designed to be similar a human's consequential communication? It can be an interesting question to ask if consequential communication from the AI to the user is even possible to design. Mutlu et al. investigated the impact of 'nonverbal leakage' in robots on human collaborators \cite{mutlu2009nonverbal}, however the leakage was designed intentionally as part of the interaction design.

In a co-creative setting, the modalities for AI initiated communication can include text, voice, visuals (icons, image, animation), haptic and embodied communication \cite{nigay2004design}. There are some communication channels that work for both human to AI and AI to human communication, such as text, voice, and embodied communication. These communication channels are under both categories to identify the possibilities based on the direction of information flow. 

\vspace{-0.2cm} 

\subsection{Interaction with the Shared Product}
Interaction components related to the shared creative product in a co-creative setting are discussed in this section and illustrated in Figure 1(b). Interaction with the shared product is divided into two subcategories, creative contribution to the product and creative process.

\subsubsection{Creative Process}
Creative process characterizes the sequence of actions that lead to a novel and creative production \cite{lubart2001models}. In COFI, there are three types of creative processes that describe the interaction with the shared product: generate, evaluate, and define. A co-creative AI can play the role of a generator, evaluator or a definer depending on the creative process. In the generation creative process, the co-creative AI generates creative ideas or artifacts. For example, a co-creative AI can generate a poem along with the user or produce music with users. Co-creative AI agents evaluate the creative contributions made by the user in a creative evaluation process. An example of creative evaluation will be analyzing and assessing a creative story generated by a user. And in a creative definition process, the AI agent will define the creative concept or explore different creative concepts along with the user. For example, a co-creative agent can define the attributes of a fictional character before a writer starts to write about the character. 

The basis of this categorization is the work of Kantosalo et al. that defines the roles of the AI as generator, evaluator, and concept definer \cite{kantosalo2016modes}. COFI adopts the categorization of Kantosalo et al. as a basis for understanding the range of potential creative processes: The generator generates artifacts in a specific conceptual description, the evaluator evaluates these concepts, and the concept definer defines the conceptual space \cite{kantosalo2016modes}. In the recent work of Kantosalo and Jordanous, they compared their defined roles with the apprentice framework of Negrete-Yankelevich's and Morales-Zaragoza, where the roles are generator, apprentice and master \cite{negrete2014apprentice}.

\subsubsection{Creative Product}
The creative product is the idea or concept that is being created. Creative product has two interaction components, contribution type and contribution similarity. We identified these specific components as we focused on various aspects of contribution making to the shared product as meaning emerges through the contributions in a collaboration. These components are identified from the literature and discussed in the following subsections. 

\textbf{Contribution Type:}
In a co-creation, an individual can contribute in different ways to the shared product. Co-creators can generate new elements for the shared product, extend the existing contribution, and modify or refine the existing contribution. How a co-creator is contributing depends on their interaction with the shared product and their interpretation of the interaction. The primary contribution types according to COFI are: 'create new', 'extend', 'transform' and 'refine'. 'Extend' refers to extending or adding on to a previous contribution made by any of the collaborators. Generating something new or creating new objects is represented by 'create new', whereas 'transform' conveys turning a contribution into something totally different. 'Refine' is evaluating and correcting a contribution with similar type of contribution. For example, in a co-creative drawing, drawing a tree will be considered 'create new'. Extend is when the collaborator adds a branch to the tree or extends the roots of the tree. Turning a tree branch into something else, such as a flower, will be considered a 'transformation', different from 'create new' as it is performed on a previous contribution to turn it into a new object. 'Refine' is when the collaborator polishes the branch of the tree to give more detail.

Contribution types are adopted and adapted from Boden’s categories of computational creativity based on different types of contribution: combinatorial, exploratory, and transformational \cite{boden1998creativity}. Combinatorial creativity involves novel (improbable) combinations of similar ideas to the existing ideas. We adapted 'expand' and 'refine' from combinatorial creativity as 'expand' is extending the existing contribution and 'refine' is about correcting or emphasizing the contribution with similar ideas. Exploratory creativity involves the generation of novel ideas by the exploration of defined conceptual spaces and 'creating new' is adapted from this as users use explores the conceptual space when creating something new. Transformational creativity involves the transformation of some dimension of the space so that new structures can be generated, which could not have arisen before and 'transform' is adapted from this.

\textbf{Contribution Similarity:}
In COFI, similarity refers to the degree of similarity or association between a new contribution compared to the contribution of the partner. Near refers to high similarity with the partner's contribution and far means less similarity with the partner's contribution. In this paper, AI agents that use ‘near’ will be referred to as pleasing agents, and agents that use ‘far’ will be referred to as provoking agents.

Miura and Hida demonstrated that high similarity and low similarity in contributions and ideas among collaborators are both essential for greater gains in creative performance \cite{miura2004synergy}. Both convergent and divergent exploration have their own value in a creative process. Divergent thinking is "thinking that moves away in diverging directions to involve a variety of aspects", whereas convergent thinking is demarcated as "thinking that brings together information focused on something specific" \cite{thefreedictionary}. Basadur et al. asserted that divergent thinking is related to the ideation phase and convergent thinking is related to the evaluation phase \cite{basadur1996measuring}. Kantosalo et al. defined pleasing and provoking AI agents, based on how similar their contributions are \cite{kantosalo2016modes}. A pleasing computational agent follows the human user and complies with the human contribution and preference. Provoking computational agents provoke the human by challenging the human-provided concepts with divergent ideas and dissimilar contribution.

\vspace{-0.0cm}  
\section{Analysis of Interaction Models in Co-creative Systems Using COFI}
\subsection{Data}
We used COFI to analyze a corpus of co-creative systems to demonstrate COFI's value in describing the interaction designs of co-creative systems. We initiated our corpus of co-creative systems using the archival website called the “Library of Mixed-Initiative Creative Interfaces” (LMICI), which archives many of the existing co-creative systems from the literature \cite{Libraryo68:online}. Mixed initiative creative systems are often used as an alternative term for co-creative systems \cite{yannakakis2014mixed}. Angie Spoto and Natalia Oleynik created this archive after a workshop on mixed-initiative creative interfaces led by Deterding et al. in 2017 \cite{deterding2017mixed, Libraryo68:online}. The archive provides the corresponding literature and other relevant information for each of the systems. LMICI archive consists of 74 co-creative systems from 1996 to 2017. However, we used 73 systems from the LMICI archive due to the lack of information regarding one system. We added 19 co-creative systems to our dataset to include recent co-creative systems (after 2017). We used the keywords 'co-creativity' and 'human-AI creative collaboration' to search for existing co-creative systems from 2017 to 2021 in the ACM digital library and Google scholar. Thus, we have 92 co-creative systems in the corpus that we used to analyze the interaction designs using COFI. Table \ref{tab:freq} shows all the co-creative systems that we analyzed with corresponding years and references. Figure \ref{} shows the count of the co-creative systems in our dataset each year.

\begin{figure}[h]
  \centering
  \includegraphics[width=0.75\linewidth]{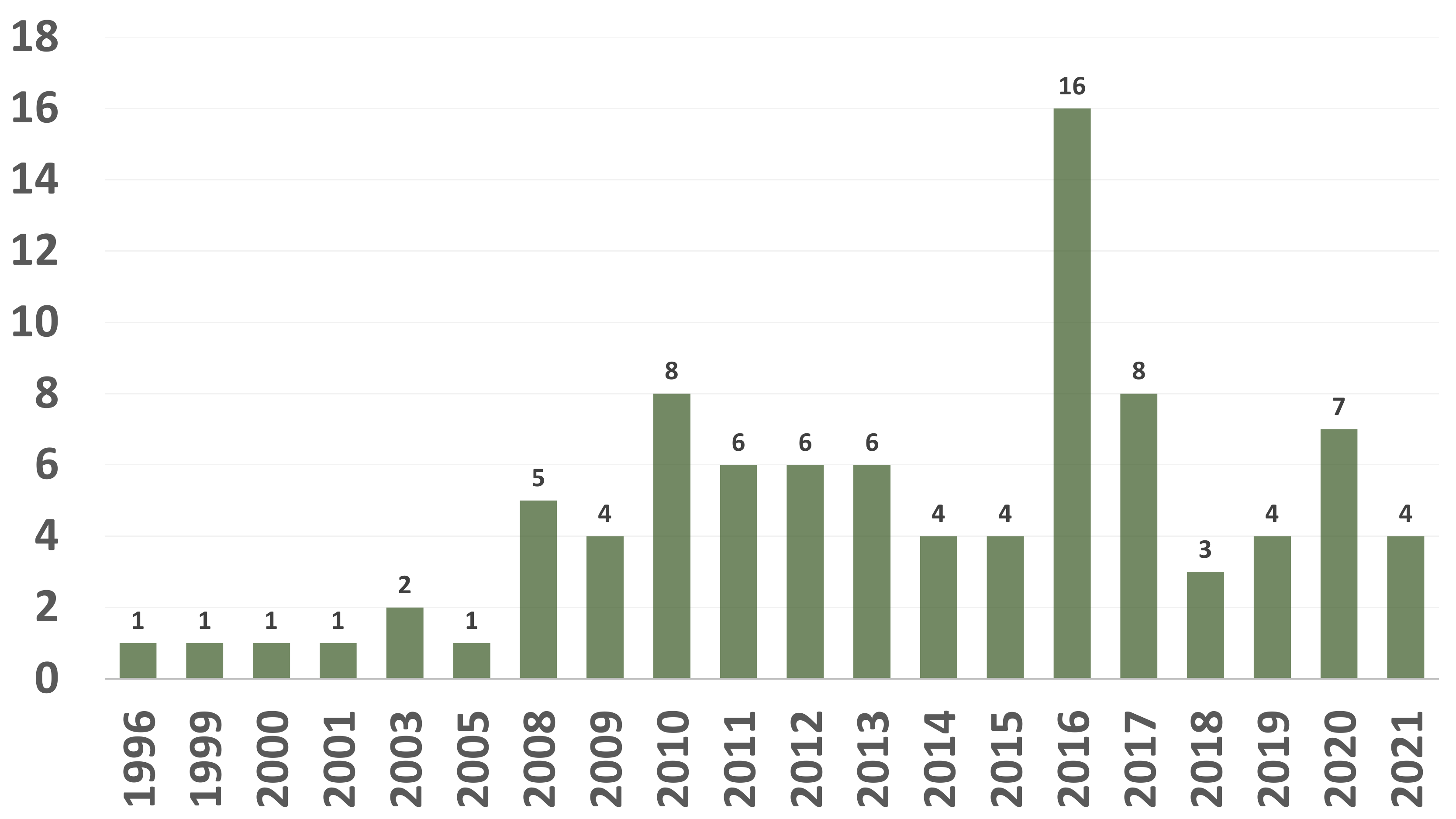}
  \caption{Counts of Co-creative Systems in the Dataset per Year.}
  \Description{}
\end{figure}

\begin{table}
  \caption{List of Co-creative Systems in the Dataset Sorted by Year.}
  \label{tab:freq}
  \begin{tabular}{cc}
    \toprule
    Year&Co-creative Systems\\
    \midrule
    1996& Improv \cite{perlin1996improv}\\
    1999& GeNotator \cite{thywissen1999genotator}\\
    2000& NEvAr \cite{machado2000nevar}\\
    2001& Metasynth \cite{dahlstedt2001mutasynth}\\
    2003& Facade \cite{mateas2003faccade}, continuator  \cite{pachet2003continuator}\\
    2005& LOGTELL \cite{ciarlini2005logic}\\
    2008& CombinFormation\cite{kerne2008combinformation},  REQUEST \cite{riedl2008toward}, miCollage\cite{xiao2008mixed}, BeatBender \cite{levisohn2008beatbender}, WEVVA \cite{nelson2017fluidic}\\
    2009& Terrain Sketching \cite{gain2009terrain}, JNETIC \cite{ bergen2009evolving}, Synthetic Audience \cite{o2011simulating}, The Poetry Machine \cite{widows_sandilands_2009}\\
    2010&  SKETCHAWORLD \cite{smelik2010interactive}, Tanagra \cite{smith2010tanagra}, Realtime Generation of Harmonic Progressions \cite{eigenfeldt2010realtime}, \\ & JamBot.   \cite{brown2010generative}, Filter Trouve \cite{colton2010experiments}, Clap-along \cite{young2010clap}, EDME \cite{lopez2010real}, LEMu \cite{lopez2010real},\\
    2011& Shimon \cite{hoffman2011interactive}, Stella \cite{leon2011stella}, Party Quirks \cite{magerko2011digital}, Generation of Tracks in a High-end Racing Game \cite{cardamone2011interactive}, \\ & ELVIRA \cite{colton2011ludic}, Creating Choreography with Interactive Evolutionary Algorithms \cite{eisenmann2011creating}\\
    2012& Spaceship Generator \cite{liapis2012co}, MaestroGenesis \cite{szerlip2012maestrogenesis}, PINTER \cite{gilroy2012exploring}, Co-PoeTryMe \cite{oliveira2014adapting}, \\ & A formal Architecture of Shared Mental Models \cite{hodhod2016closing}, Impro-Visor \cite{keller2012continuous}\\
    2013& Sentient Sketchbook \cite{liapis2013sentient}, Dysphagia \cite{shaker2013ropossum}, Viewpoints AI \cite{jacob2013viewpoints}, \\ & Ropossum \cite{dipaola2013adaptation}, COCO Sketch\cite{davis2013human}, Sentient World\cite{liapis2013sentient}\\
    2014& Chef Watson \cite{pinel2014computational}, Kill the Dragon and Rescue the Princess \cite{laclaustra2014kill}, Nehovah \cite{smith2014nehovah}, Autodesk Dreamcatcher\cite{oliveira2014adapting} \\
    2015& CAHOOTS \cite{wen2015omg}, Funky Ikebana \cite{compton2015casual}, StyleMachine \cite{metacreative_technologies_2015}, Drawing Apprentice \cite{davis2015drawing}\\
    2016& Improvised Ensemble Music Making on Touch Screen \cite{martin2016intelligent}, AceTalk \cite{trinh2016thinking}, Chor-rnn \cite{crnkovic2016generative}, Cochoreo \cite{carlson2016cochoreo}, \\ & Evolutionary Procedural 2D Map Generation \cite{scheibenpflug2016evolutionary}, Danesh \cite{cook2016danesh}, Plecto \cite{ianigro2016plecto}, \\ & Image-to-Image \cite{isola2017image}, Robodanza \cite{infantino2016robodanza}, SpeakeSystem \cite{yee2016experience}, TaleBox\cite{castano2016talebox}, \\ &  ChordRipple \cite{samuel2016design},  Robovie \cite{kahn2016human}, Creative Assistant for Harmonic Blending \cite{kaliakatsos2016argument}, \\ & Writing Buddy \cite{samuel2016design}, Recommender for Game Mechanics \cite{machado2016shopping}\\
    2017& TOPOSKETCH \cite{white2017generating}, Trussfab\cite{kovacs2017trussfab}, Chimney \cite{morreale2017renegotiating}, FabMachine \cite{kim2017machines}, \\ & LuminAI \cite{long2017designing}, GAIA \cite{goel2017gaia}, 3Buddy \cite{lucas2017stay}, Deeptingle \cite{khalifa2017deeptingle}\\
    2018& The Image Artist \cite{zoric2018image}, DuetDraw \cite{oh2018lead}, Robocinni \cite{ackerman2018co}\\
    2019& In a silent way \cite{mccormack2019silent}, Metaphoria \cite{gero2019metaphoria}, collabDraw \cite{fan2019collabdraw}, DrawMyPhoto \cite{williford2019drawmyphoto}\\
    2020& Shimon the Rapper \cite{savery2020shimon}, ALYSIA \cite{cheatley2020co}, Cobbie \cite{lin2020your}, WeMonet \cite{li2020empowering}, \\ & Co-cuild \cite{deshpande2020towards}, IEC \cite{parente2020type}, Creative Sketching Partner \cite{karimi2020creative}\\
    2021& BunCho \cite{osone2021buncho}, CharacterChat \cite{schmitt2021characterchat}, StoryDrawer \cite{zhang2021storydrawer}, FashionQ \cite{jeon2021fashionq}\\

  \bottomrule
\end{tabular}
\end{table}

\begin{figure}[h]
  \centering
  \includegraphics[width=0.95\linewidth]{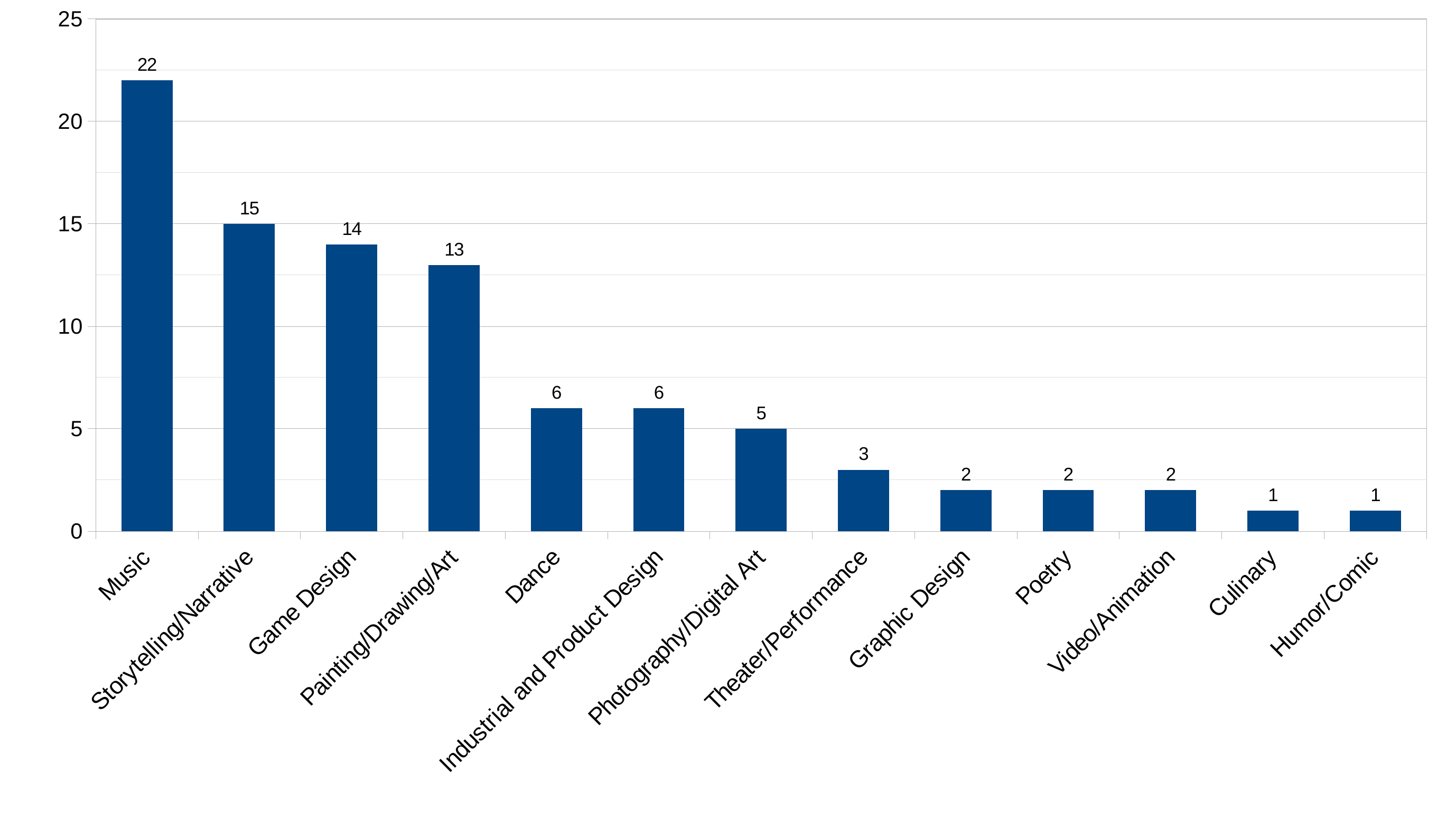}
  \caption{Count of Co-Creative Systems in Different Creative Domains.}
  \Description{}
\end{figure}

We grouped the systems into 13 categories describing their creative domains. The categories are Painting/Drawing/Art, Culinary, Dance, Music, Storytelling/Narrative/Writing, Game Design, Theatre/Performance, Video/Animation, Photography, Poetry, Industrial and Product Design, Graphic Design and Humor/Comic. In Figure 4, the count of the systems in each category is provided. We see the most common creative domains in the corpus are music, storytelling/narrative/writing, Game design and Painting/Drawing/art. The distribution shows that some creative domains are are not well represented in this dataset or rarely used in developing co-creative systems, for example, culinary, humor, and graphic design.

\subsection{Coding Scheme}
To analyze the interaction design of the existing co-creative systems, we coded the interaction designs of 92 systems using COFI. Two coders from our research team independently coded 25\% of the systems following COFI. They then achieved consensus through discussing the disagreements in the codes (Kappa Inter-rater reliability  0.79). The rest of the systems were coded by a single coder according to the consensus. For each system, the coding shows all interaction design components according to COFI. All the interaction components of the systems were coded according to the information provided in the corresponding literature. For a specific interaction component, when none of the subcategories are present in the interaction design, we coded it as ‘None’.  
\vspace{-0.2cm} 

\subsection{Interaction Design Models among Co-creative Systems}
For identifying different interaction models utilized by the co-creative systems in the dataset, we clustered all the systems using their interaction components. We used K-modes clustering \cite{huang1997clustering, cao2012dissimilarity} for identifying clusters as the K-modes algorithm is suitable for categorical data. K-modes clustering is an extension of K-means, but instead of means, this algorithm uses modes. For demonstrating the cluster centroids, this algorithm uses modes of all the features. We used all the interaction components according to COFI as features. We found three clusters of the systems based on their interaction design (Figure 5). The first cluster includes 67 co-creative systems and thus indicating a dominant interaction model. The second cluster includes 9 systems and the third one includes 16 systems. We used chi-square for determining interaction components that contribute significantly to forming the clusters and found that all of the interaction components are significant factors for the clusters (all P values < 0.05). Figure 5 shows the three major interaction models, including all the interaction components (cluster centroids represented by feature modes). 

\begin{figure}[h]

  \centering
  \includegraphics[width=1\linewidth]{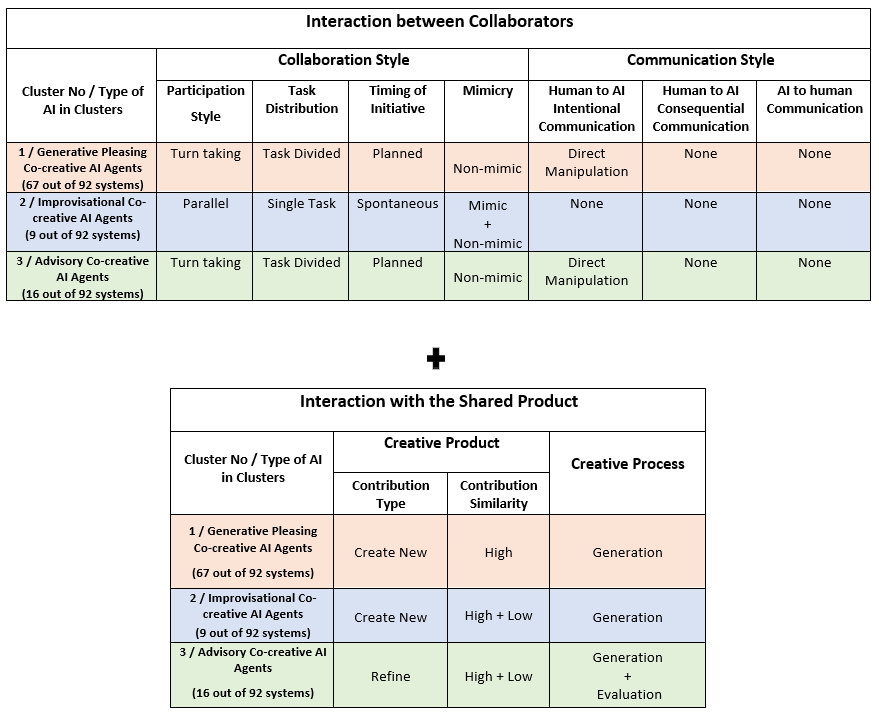}
  \caption{Interaction Designs for the Three Clusters of Co-creative Systems.}
  \Description{}
\vspace{-0.6cm} 
\end{figure}

\subsubsection{Cluster 1 - Interaction Design Model for Generative Pleasing AI Agents}
The interaction model of the first cluster is the most prevalent as there are 67 systems in this cluster sharing the same or similar model. This dominant interaction model shows that most of the co-creative systems in the dataset utilize turn-taking as the participation style. Therefore, each of the collaborators must wait until the partner finishes their turn. This interaction model uses 'planned' timing of initiative which is an indication of non-improvisational co-creativity. Hence, most of the systems in the dataset do not support improvisational creativity. This interaction model uses direct manipulation for human to AI intentional communication. However, this model does not incorporate any human to AI consequential communication or AI to human communication. The main task is divided between the collaborators, and the AI agent uses generation as the creative process in most of the systems in this cluster and creates something new without mimicking the user. The degree of similarity in contribution is high. In other words, the AI agent pleases the human by generating contributions that follow along with the contributions made by the human. Mostly, this interaction model is used by non-improvisational systems that generate creative products to please the users.

An example of a system that uses this interaction design is Emotion Driven Music Engine (EDME) \cite{lopez2010real}. EDME generates music based on the emotions of the user. The user selects an emotion, and EDME plays music to match that emotion. This system works in a turn-taking way with the user. The timing of initiative-taking is planned as the system will always respond after the human finishes selecting their emotion. The task is divided between the collaborators as the user defines the conceptual space by choosing an emotion from the interface and the system generates the music according to that emotion. The system contributes to the collaboration by creating something new and without mimicking the user. The system creates music that is associated with and similar to the user-defined emotion. The biggest challenge here is the human can not give any feedback or communicate with the system regarding the generated music. The system can not track any consequential information from the human such as facial expression, eye gaze and embodied gestures. Also, the system can not communicate any relevant information to the user such as providing additional information regarding the contribution or visual cues. 

\subsubsection{Cluster 2 - Interaction Design Model for Improvisational AI Agents}
\vspace{-0.1cm}
The interaction design for the systems in cluster 2 uses parallel participation style where both agents can contribute simultaneously. The task distribution for these systems is usually ‘same task’ and most of the systems contribute by generating in the creative process. Most of the systems in this cluster contribute to the collaboration by creating something new and these systems can do both mimicry and non-mimicry. The degree of similarity in terms of users’ contribution can be both high and low. This interaction model employs spontaneous initiative-taking while both co-creators contribute to the same task with parallel participation style, indicating improvisational co-creativity. Systems in this cluster do not have any way of communication between the user and the system, and a lack of communication in improvisational co-creativity can reduce the collaboration quality and engagement \cite{hoffman2011interactive}.

An example system for this cluster is LuminAI, where human users improvise with virtual AI agents in real time to create a dance performance \cite{long2017designing}. Users move their body and the AI agent will respond with an improvised movement of its own. Both the AI agent and users can dance simultaneously and they take initiatives spontaneously. Collaborators contribute to only a single task, generating dance movements. The AI can create new movements and transform user movements while it can do both mimicry and non-mimicry. The dance movements can be similar or different from the user. There is no way the user can deliberately communicate with the system or the system can communicate with the user. Here, the creative product itself is an embodied product but the system can not collect any consequential information from the user such as eye gaze, facial expression or additional gestures other than dance moves.

\subsubsection{Cluster 3 - Interaction Design Model for Advisory AI Agents}
\vspace{-0.1cm}
The third cluster includes systems that work in a turn taking manner and the task is divided into subtasks between the collaborators. The initiative taking is planned prior to the collaboration. Users can communicate to the system through direct manipulation, but there is no human to AI consequential communication channel or AI to human communication channel. The most notable attribute for this interaction model is both the generation and evaluation ability of the AI agent unlike the other two interaction models where the AI agent can only contribute by generating. Systems with this interaction model can act as an adviser to the user by evaluating the contribution of the user. Most of these systems in this cluster contribute by refining the contribution of the user. These systems do not mimic the contribution of the user and the degree of contribution similarity can be both high and low.

An example of co-creative systems that utilize this model is Sentient World which assists video game designers in creating maps \cite{liapis2013sentient}. The designer creates a rough terrain sketch, and Sentient World evaluates the map created by the designer and then generates several refined maps as suggestions. This system works in a turn-taking manner with the user, and the initiative taking is planned. The AI agent uses both generation and evaluation as creative processes by generating maps and evaluating maps created by the user. The user can communicate with the system minimally with direct manipulation (clicking buttons) for providing user preference for the maps. The AI agent can not communicate any explicit information to the human and can not collect any consequential information from the user such as facial expression, eye gaze and embodied information. Sentient World can both create new maps and refine the map created by the user. The system does not mimic the user contribution and the similarity with user contribution is high.
\vspace{-0.2cm} 

\subsection{Adoption Rate of the Interaction Components used in the Systems}
Figure 6 shows the adoption rate of each of the interaction components in COFI used in the systems. The first section of the table comprises interaction components under \emph{collaboration style}. Turn-taking is the most common participation style in the dataset (89.1\%), while just 10.9\% of the systems use parallel participation. Parallel participation is used by the systems that engage in performative co-creation. Most of the co-creative systems in the dataset use task-divided distribution of tasks (75\%) as they work on separate creative subtasks. 25\% systems use \emph{same task} as their task distribution as both the user and the AI work on the same creative task/s. Timing of initiative is planned in 86.8\% of the systems and the rest of the systems take spontaneous initiatives without any fixed plan. For mimicry, 90.2\% of the systems employ non-mimicry, 8.7\% systems use both mimicry and non-mimicry, and only one system (1.1\%) uses mimicry.

The second category, \emph{communication style}, is concerned with different communication channels used by the co-creative systems. 69.6\% systems use direct manipulation as the human to AI communication channel. Voice, embodied and text is used rarely by the systems. 3.3\% of the systems use embodied communication as human to AI consequential communication and most of the systems (95.7\%) do not track and collect any consequential information from the user. For AI to human communication, most systems do not have any channels. In the next section, we talk about the trend in communication channels in co-creative systems. 

In the creative process category, it is noticeable that the majority of the systems (79.3\%) employ generation as the creative process and 15.2\% of the systems use both generation and evaluation as the creative processes. Definition as a creative process is rarely used in the co-creative systems. 

In the creative product category, contribution type is the first interaction component and most co-creative systems use \emph{create new} (59.8\%). 10.9\% of the systems use both \emph{create new} and \emph{refine} as the contribution type. 8.7\% of the systems use both \emph{create new} and \emph{extend} as the contribution type.

\begin{figure}[h]
  \centering
  \includegraphics[width=1\linewidth]{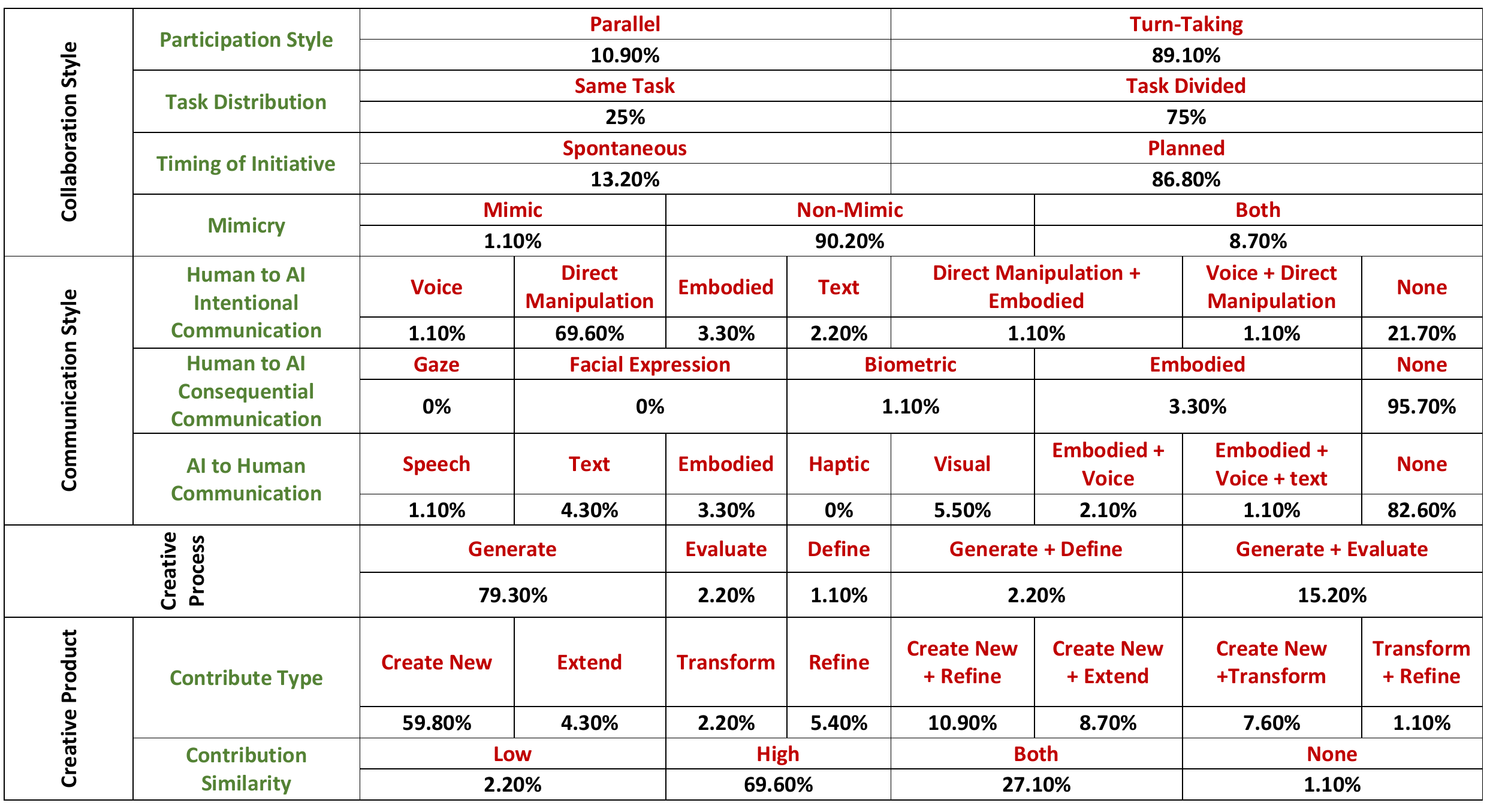}
  \vspace{-0.2cm}  
  \caption{Adoption Rate of Each Interaction Component used in the Co-creative Systems in the Dataset.}
    \vspace{-0.3cm}  
  \Description{}

\end{figure}

\subsection{Communication in Interaction Models}
Our analysis identifies a significant gap in the use of the components of interaction in the co-creative systems in this dataset: a lack of communication channels between humans and AI (Figure 3). In co-creative systems, subtle communication happens during the creative process through contributions. For example, in a collaborative drawing co-creative system where no communication channel exists between the user and the AI, subtle interaction happens through the shared product as co-creators make sense of each other's contribution and then make a new contribution. Designing different modalities for communication between the user and the AI has the potential to improve the coordination and quality of collaboration. However, 82.6\% of the systems cannot communicate any feedback or information directly to the human collaborator other than communicating through the shared product. The rest of the systems communicate with the users through text, embodied communication, voice, or visuals (image and animation). For human to AI consequential communication, 95.7\% of the systems can not capture any consequential information from the human user such as facial expression, biometric data, gaze and postures. However, consequential communication can increase user engagement in collaboration. For the intentional communication from human to AI, most of the systems use direct manipulation (clicking buttons or selecting options) to communicate (69.6\%). In other words, in most of the systems, users can only minimally communicate with the AI or provide instructions to the AI directly, for example, through clicking buttons or using sliders. 21.7\% of the systems have no way for the user to communicate with the AI intentionally. The rest of the systems use other intentional communication methods, like embodied communication or voice or text.
\begin{figure}[h]
  \centering
  \includegraphics[width=0.9\linewidth]{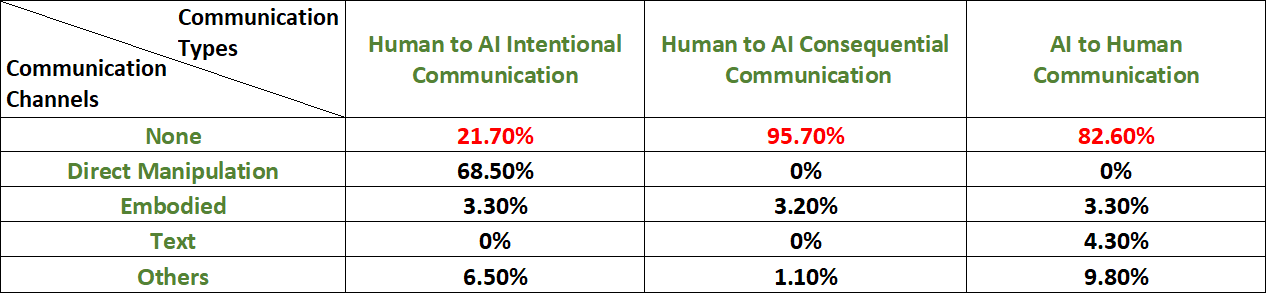}
  \vspace{-0.2cm}  
  \caption{Distribution of Different Kinds of Communication between Humans and AI in the Co-Creative Systems in the Dataset.}
    \vspace{-0.3cm}  
  \Description{}

\end{figure}

Some of the systems in our dataset utilize multiple communication channels. Shimon is a robot that plays the marimba alongside a human musician \cite{hoffman2011interactive}. Using embodied gestures as visual cues to anticipate each other’s musical input, Shimon and the musician play an improvised song, responding to each other in real-time. The robot and the human both use intentional embodied gestures as visual cues to communicate turn-taking and musical beats. Therefore, this system includes human to AI intentional communication and AI to human communication. Findings from a user study using Shimon demonstrate that visual cues aid synchronization during improvisational co-creativity. Another system with interesting communication channels is Robodanza, a humanoid robot that dances with humans \cite{infantino2016robodanza}. Human dancers use intentional communication by intentionally touching the robot’s head in order to awaken it and the robot tracks human faces to detect consequential information. The robot is able to detect the noise and rhythm of hands clapping and tapping on a table. The robot can move its head in the direction of the perceived rhythms and move its hand following the perceived tempo for communicating its status to the human users.

\section{Discussion}
We develop and describe COFI to provide a framework for analyzing, comparing, and designing interaction in co-creative systems. Researchers can use COFI to explore the possible spaces of interaction for choosing an appropriate interaction design for a specific system. COFI can be beneficial while investigating and interpreting the interaction design of existing co-creative systems. As a framework, COFI is expandable as other interaction components are added in the future. We analyzed the interaction models of 92 existing co-creative systems using COFI to demonstrate its value in investigating the trends and gaps in the existing interaction designs in co-creative systems. We identified three major clusters of interaction models utilized by these systems. In the following paragraphs, we explain the interaction models and discuss the potential for further research in specific interaction components. These interaction models can be useful when designing a co-creative system since they can help identify appropriate interaction components and determine if interaction components should be modified for the corresponding type of co-creative AI agent.

The most common interaction model in our dataset is suitable for generative co-creative AI agents that follow and comply with human contributions and ideas by generating similar contributions. Provoking agents are rare in the literature, and in fact, such a stance seems to be opposed by some in the literature. For example, Tanagra's creators ensured "that Tanagra does not push its own agenda on the designer" \cite{smith2010tanagra}. However, both pleasing and provoking agents have use-cases within co-creative systems \cite{kantosalo2016modes}. For example, if a user is trying to produce concepts or ideas that convey their specific style, a pleasing agent that contributes similar ideas is more desirable. However, if a user is searching for varied ideas, a provoking agent with different contributions is an ideal creative partner as it will provide more divergent ideas. This model can be improved with consequential communication tracking from users and AI to human communication.

The second interaction model is suitable for improvisational AI agents as it uses spontaneous initiative-taking and both agents work on the same task in parallel. Additionally, this model includes both mimicry and non-mimicry, unlike the other models which direct the AI to take proper action in an improvisational performance. This model can be utilized as a guide while designing interaction in an improvisational co-creative system. However, this model does not include any intentional or consequential communication channels from humans to AI or AI to humans, which can negatively impact the collaboration quality and user experience, especially in improvisational co-creativity where communication is the key. Hoffman et al. asserted that communication aids synchronization and coordination in improvisational co-creativity \cite{hoffman2011interactive}. Further research can extend this model by including or extending human-AI communication channels.

The third interaction model is used by co-creative AI agents that work as an advisor by evaluating user's contributions and contributing to the shared product as a generator. In product-based co-creation, AI agents that can both generate and evaluate help the user generate precise creative ideas and artifacts. For example, in industrial design, the co-creative AI agent can help in creative ideation by evaluating the user-provided concept for a robust and error-free design and also help in the generation of the artifact with divergent or convergent ideas \cite{kovacs2017trussfab}. AI agents that use this model can refine the user's contributions in contrast to the other models. The limitations of this model include the absence of human to AI consequential communication and AI to human communication.

A notable finding from the analysis of this dataset is the lack of AI agents defining the conceptual space as the creative process (only 4 out of 92). Most of the systems in the corpus contribute by generating and some contribute by evaluating the human contributions. In the context of co-creativity, defining the conceptual space is an essential task. An AI agent can define the conceptual space without any guidance from the user. For example, the Poetry Machine is a poetry generator that prompts the user with images that users respond to with a line of poetry \cite{Libraryo68:online, widows_sandilands_2009} and then organizes the lines of poetry into a poem. An AI agent can also suggest multiple ideas for the conceptual space while the user can select their preferred one. TopoSketch \cite{white2017generating} generates animations based on a photo of a face provided by the human and displays various facial expressions as ideas for the final animation. CharacterChat inspires writers to create fictional characters through a conversation. The bot converses with the user to guide the user to define different attributes of the fictional character. Humans may desire inspiration for creative concepts and ideas at the beginning of a creative journey. Creative brainstorming and defining creative concepts can be potential research areas for co-creative systems. There is potential for designing new co-creative systems that both define the creative conceptual space and explore it with the user. 

The most significant area of improvement in all of the interaction models identified is communication, the key to coordination between two agents. Providing feedback, instructions or conveying information about the contribution is essential for creative collaboration. Without any communication channel between the co-creators, the creation becomes a silent game \cite{sorensen2016silent, sorensen2017exploring} as collaborators can not express any concerns and provide feedback about their contributions. Communication through the creative product is subtle communication and may not be enough to maintain the coordination and collaboration quality. Most of the existing co-creative systems in our dataset have minimal communication channels, and this hinders the collaboration ability of the AI agent and the interactive experience. Most of the systems in the dataset utilize only direct manipulation for communicating intentional information from the users. Direct manipulations include clicking buttons and using sliders for rating AI contribution, providing simple instructions and collecting user preferences. For most systems, direct manipulation provides a way for minimal communication and does not provide users with a way to communicate more broadly. Very few systems in the dataset use other communication channels other than direct manipulation for human to AI intentional communication. For example, AFAOSMM (2012) \cite{hodhod2016closing} is a theatre-based system that uses gestures as intentional communication and Robodanza (2016) \cite{infantino2016robodanza} uses embodied movements along with direct manipulation for intentional communication. Human to AI consequential communication is rarely used in the systems but an effective way to improve creative collaboration. It has been demonstrated that humans, during an interaction, can reason about others’ ideas, goals, intentions and predict partners’ behaviors, a capability called Theory of Mind (ToM) \cite{premack1978does, yoshida2008game, baker2017rational}. Having a Theory of Mind allows us to infer the mental states of others that are not directly observable, enabling us to engage in daily interaction. The ability to intuit what others think or want from brief nonverbal interactions is crucial to our social lives as we see others' behavior not just as motions but as an intentional action. In a collaboration, Theory of Mind is essential to observe and interpret the behavior of a partner, maintain coordination and act accordingly. Collecting unintentional information from the human partner has the potential to improve the collaboration and user experience in a human-AI co-creation, and may lead to enabling AI to mimic the Theory of Mind ability of humans. The technology for collecting consequential information from the user includes eye trackers, facial expression trackers, gesture recognition devices, and cognitive signal tracking devices. 

AI to human communication channels are also rarely utilized in the identified interaction models. However, it is essential to understand the AI partner by the users to build an engaging and trustworthy partnership. Many intelligent systems lack the core interaction design principles such as transparency and explainability and it makes them hard to understand and use \cite{eiband2021support}. To address the challenge of transparency of AI interaction should be designed to support users in understanding and dealing with intelligent systems despite their complex black-box nature. When AI can communicate its decision making process to users and explain its contribution, the system becomes more comprehensible and transparent to build a partnership. So, AI to human communication is critical for interaction design in co-creative systems. Visuals, text, voice, embodied, and haptic feedback can be used to convey information, suggestions, and feedback to the users. There is a distinction between AI to human communication and AI steerability. For example, LuminAI is a co-creative AI that dances with humans \cite{long2017designing}. Here the generated creative product is dance, an embodied product created by gestures and embodied movements. However, AI can only communicate by contributing to the product and does not directly communicate to humans. Humans can steer the AI by contributing different embodied contributions to the final product and the AI generates contributions based on the user movements. This is different from embodied communication that intentionally communicates that the collaborator is doing great with a thumbs up. The gap in interaction design in terms of communication is an area of future research for the field of co-creativity. User experiments with different interaction models can help identify effective interaction design for different types of co-creative systems \cite{rezwanaa2021creative}. COFI provides a common framework for analyzing the interaction designs in existing co-creative systems to identify trends and gaps in existing interaction designs for designing improved interaction in a co-creative system. 

AI is being used increasingly in collaborative spaces, for example, recommender systems, self-driving vehicles, and health care. Much AI research has focused on improving the intelligence or ability of agents and algorithms \cite{ashktorab2020human}. As AI technology shifts from computers to everyday devices, AI needs social understanding and cooperative intelligence to integrate into society and our daily lives. AI is, however, a novice when it comes to collaborating with humans \cite{dafoe2021cooperative}. The term 'human-AI collaboration' has emerged in recent work studying user interaction with AI systems \cite{wang2019human, oh2018lead, cai2019hello, arous2020opencrowd}. This marks both a shift to a collaborative from an automated perspective of AI, and the advancement of AI capabilities to be a collaborative partner in some domains. Ashktorab et al. asserted that human-AI co-creation could be a starting point of designing and developing AI that can cooperate with humans \cite{ashktorab2020human}. Human-AI interaction has many challenges and is difficult to design \cite{yang2020re}. HCI deals with complex technologies, including research to mitigate unexpected consequences. A critical first step in designing valuable human-AI interactions is to identify technical challenges, articulate the unique qualities of AI that make it difficult to design, and then develop insights for future research \cite{yang2020re}. Building a fair and effective AI application is considered difficult due to the complexity both in defining the goals and algorithmically achieving the defined goals. Prior research has addressed these challenges by promoting interaction design guidelines \cite{mitchell2019model, amershi2019guidelines}. In this paper, we provide COFI as a framework to describe the possible interaction spaces in human-AI creative collaboration and identify existing trends and gaps in existing interaction designs. COFI can also be useful in AI research and HCI research to design cooperative AI in different domains. COFI will expand as we learn and identify more aspects of human-AI collaboration. 

\vspace{-0.3cm} 

\section{Limitations}
The identification of clusters of interaction models in human-AI co-creative systems is limited to the specific dataset that we used for the analysis. Although we believe this sample contains a large population, the systems in the dataset are limited by the expectations and technologies at the time of publication. We expect the clusters and descriptions of interaction models for co-creative systems will change over time. 

\section{Conclusions}
This paper develops and describes the COFI as a framework for modeling interaction in co-creative systems. COFI was used to analyze the interaction design of 92 co-creative systems from the literature. Three interaction models for co-creative systems were identified: generative pleasing agents,  improvisational agents, and advisory agents. When developing a co-creative system, these interaction models can be useful to choose suitable interaction components for corresponding co-creative systems. COFI is broader than the interaction designs utilized in any specific co-creative system in the data set. The findings show that the space of possibilities is underutilized. While the analysis is limited to the data set, it demonstrates that COFI can be a tool for identifying research directions and research gaps in the current space of co-creativity. COFI revealed a general lack of communication in co-creative systems within the dataset. In particular, very few systems incorporate AI to human communication, communication channels other than direct manipulation for collecting intentional information from humans and gathering consequential communication data, such as eye gaze, biometric data, gesture, and emotion. This gap demonstrates an area of future research for the field of co-creativity. We argue that COFI will provide useful guidelines for interaction modeling while developing co-creative systems. As a framework, COFI is expandable as other interaction components can be added to it in the future. User experiments with different interaction models can help identify effective interaction design for different types of co-creative systems and lead to insights into factors that affect user engagement.
\bibliographystyle{ACM-Reference-Format}
\bibliography{sample-base}

\end{document}